\documentclass[runningheads]{llncs}
\usepackage{amsmath}
\usepackage{amssymb}
\usepackage{txfonts}
\usepackage{mathdots}
\usepackage{graphicx}
\usepackage{tabularx}
\usepackage{multirow}
\usepackage{subcaption}
\usepackage{booktabs}
\usepackage{xcolor}
\usepackage{graphicx}
\usepackage{hyperref}
\usepackage{comment}
\newcommand{\tmpframe}[1]{\fbox{#1}}
\begin{document}
%

\title{PiXi: Password Inspiration by Exploring Information}
%
%


\author{Shengqian Wang \and
Amirali Salehi-Abari \and
Julie Thorpe}
%
\authorrunning{S. Wang et al.}
%
\institute{Ontario Tech University, Oshawa, Canada} 
\maketitle              

\begin{abstract}
Passwords, a first line of defense against unauthorized access, must be secure and memorable. However, people often struggle to create secure passwords they can recall. To address this problem, we design \textit{\textbf{P}assword \textbf{i}nspiration by e\textbf{X}ploring \textbf{i}nfor\-mation (PiXi)\footnote{This preprint has not undergone peer review or any post-submission improvements or corrections. The Version of Record of this contribution has been accepted to appear in ICICS 2023.}}, a novel approach to nudge users towards creating secure passwords. PiXi is the first of its kind that employs a password creation nudge to support users in the task of generating a unique secure password themselves. PiXi prompts users to explore unusual information right before creating a password, to shake them out of their typical habits and thought processes, and to inspire them to create unique (and therefore stronger) passwords. PiXi's design aims to create an engaging, interactive, and effective nudge to improve secure password creation.
We conducted a user study ($N=238$) to compare the efficacy of PiXi to typical password creation. 
Our findings indicate that PiXi's nudges do influence users' password choices such that passwords are significantly longer and more secure (less predictable and guessable).


\keywords{Passwords \and Authentication \and Nudging \and  User Studies}
\end{abstract}

\section{Introduction}
Despite decades of development in password authentication alternatives, the majority of websites still require passwords for authentication. Unfortunately, due to time constraints, labor costs, lack of expertise, or apathy, a significant number of people reuse passwords or choose simple, predictable passwords (e.g., birthdays or names). These insecure password choices don't imply users' lack of intelligence or motivation, but may simply be due to their lack of inspiration or guidance when confronted with a blank password field. Frustration can also arise from unhelpful password policy suggestions, such as ``please use special characters to make your password stronger'' or ``make your password longer to create a strong password.'' Unfortunately, few solutions exist to support users with creating secure passwords in such helpless situations. 


While password managers, when used with random password generators, can improve password security \cite{180204,Password_Meter_1,281272}, some users are not comfortable using them. Even some organizations (e.g., governments) do not recommend their use for sensitive accounts due to the fear of the password manager vault being compromised \cite{CanadianGovt}. Password managers still require a strong master password as the key to encrypt the stored passwords in the vault. 
Therefore, regardless of employing password managers or not, users still require support for creating secure and memorable passwords for (at least) these sensitive accounts.

Nudging is a promising technique that can encourage users to create more secure and memorable passwords. However, most nudges in password systems apply a one-size-fits-all approach and primarily focus on password meters \cite{peer2020nudge,acquisti2017nudges,180204}, which use rigorous password standards to convince users to adjust their passwords to satisfy specific requirements. 
Unfortunately, many users find effective password meter designs to be annoying \cite{180204}.

To address these shortcomings, we design \textit{\textbf{P}assword \textbf{i}nspiration by e\textbf{X}ploring \textbf{i}nfor\-mation (PiXi)}, 
a novel approach to nudge users towards creating secure passwords. PiXi is the first of its kind that employs a password creation nudge to support users in the task of generating a unique password themselves. PiXi prompts users to explore unusual information right before creating a password, to shake them out of their typical habits and thought processes, and to inspire them to create unique (and therefore stronger) passwords.
We implemented and evaluated a web-based version of PiXi to answer our research questions: (Q1) Which nudges in PiXi are most effective, and do they influence users' password choices?  (Q2) Does our PiXi system support users to create more secure passwords?  
(Q3) How usable is our PiXi system, and how can its usability be improved?

%
%
%
%
%
%
%
%

To investigate these research questions, we conducted a user study ($N=238$) to evaluate the security and usability of passwords generated by users of PiXi.   
Our contributions and findings include: (i) The design of PiXi---a novel approach to nudging users to create secure passwords. (ii) Security analysis of passwords produced with PiXi.  Our study results indicate that PiXi successfully influences users' password choices, such that passwords are longer and more secure (less guessable) than a control group using a typical password creation process. (iii) Usability analysis of the PiXi system.  Our study results indicate that PiXi shows promising usability in terms of user perception and memorability. (iv) Analysis of nudge efficacy of PiXi.  Our findings indicate that some nudges are more effective than others and that PiXi's combination of nudges do influence users' password choices. 

\section{Related Work} \label{bg}
We first introduce nudging in its most general form, then highlight some of its key applications. We then narrow down our focus to nudges at the time of password creation for graphical passwords and text passwords. Finally, we summarize the key differences between our approach with others.

\vskip 2mm
\noindent \textbf{Nudging.} Nudging is a promising strategy to alter people's behavior without limiting their choices or economic incentives \cite{thaler2008nudge}. Nudges can successfully change people's decisions by minor and inexpensive interventions \cite{dijksterhuis2000relation}.   Nudging has been applied in a variety of domains including education \cite{breman2011give}, ethics \cite{bazerman2012behavioral}, social context \cite{johnson2003defaults}, health \cite{milkman2011using},  finance \cite{cai2020nudging,thaler2004save}, energy savings \cite{costa2013energy}, privacy \cite{acquisti2017nudges}, and security \cite{collier2018nudge}.  Computer security experts and administrators have recently been investigating nudges to encourage secure behaviors (see this survey for a great overview \cite{zimmermann2021nudge}).



\vskip 2mm
\noindent \textbf{Password Creation Nudges.}
Nudging techniques have been employed, with varying degrees of success, to enhance the security of both graphical and text passwords. Throughout this review, we describe each nudge using the nudge categorization of Caraban et al. \cite{caraban201923}.

\vskip 1mm
\noindent \textit{Nudges in Graphical Passwords.}
Graphical passwords are a type of knowledge-based authentication that involves remembering (parts of) images instead of a word. Some notable examples of graphical passwords and their variants are  Draw-A-Secret (DAS) \cite{DAS}, PassPoints \cite{PassPoints}, CCP \cite{CCP}, and GeoPass \cite{GeoPass,thorpe2007human}), PassFaces \cite{passfaces}, and VIP \cite{bicakci2009graphical}).  Background Draw-A-Secret (BDAS)\cite{BDAS} arguably is the first attempt to nudge users away from typical patterns during graphical password selection. It presents users with a background image, on which they need to draw their graphical password. Its background image evokes the ``salience bias", thus facilitating the creation of different graphical passwords than if the background image was not present. Zezschwitz et al. \cite{vonZezschwitz2016} used similar nudging techniques to help users create stronger patterns on Android mobile devices. Persuasive Cued Click Points (PCCP) \cite{PCCP} can be considered a facilitate (suggesting alternatives) nudge where users have to select from a point within a randomly positioned view-port (all other options are not available). Some PassPoints variations (e.g.,  \cite{ThorpePriming,PARISH2021102913,Katsini}) can be considered to employ both facilitate (hiding) and reinforce (subliminal priming) nudges. They aim to nudge users away from common patterns by presenting the background image differently at password creation for each user \cite{ThorpePriming,PARISH2021102913}. Since these nudges temporarily hide certain options (making them harder to reach), they can be categorized as facilitating (hiding) nudges.  


\vskip 1mm
\noindent \textit{Nudges in Text Passwords.}
The most straightforward way to nudge strong password selection is to suggest a random password \cite{NasrullahSystemAssigned} to the user. This is a form of facilitate (default) nudge if implemented so the user has a choice to accept the random password or not. However, memorability is a significant problem for system-assigned random passwords \cite{1341406}. Password managers can help users remember a random password without increasing memory burden \cite{Das2014TheTW}, but many users still hesitate to adopt them \cite{281272,PearmanPM,StobertPM}, due to lack of awareness and knowledge \cite{AlbayramPMAW,Seiler-HwangPMTrust}, concerned about the trust \cite{AlodhyaniPMTrust,Seiler-HwangPMTrust} or security relevant \cite{PearmanPM,alkaldi2016PM}. Even for those users who are successfully nudged to choose a random password and store it in a password manager, it is recommended to avoid using password managers for sensitive accounts (e.g., email, financial, workplace, etc.) \cite{CanadianGovt}. For these reasons, finding other ways to nudge users towards creating secure passwords remains of interest. One way to nudge users towards creating stronger text passwords is through password meters \cite{180204}. Employing a confront (friction) nudge, they provide real-time feedback on password strength to motivate users to revise their passwords, but has limited effectiveness on not ``important'' accounts \cite{EgelmanPassMeter}. Other approaches employ a facilitate (suggesting alternatives) nudge that suggests modifications to the initial password to make it secure \cite{forget2008improving,Houshmand}. However, these systems are often vulnerable to Guided Brute Force attacks \cite{schmidt_acsac13}.

\vskip 2mm
\noindent \textbf{Our Work vs Others.} The existing approaches to nudge stronger text passwords are either (a) default nudges to use a randomly generated password (typically employed as a nudge in password managers \cite{281272}), (b) confront (friction) nudges that aim to increase user's awareness of their chosen password's weakness, with no facilitation in coming up with a new password (e.g., password meters \cite{ur2015measuring}), and (c) facilitate (suggesting alternatives) nudges that suggest modifications to a user's initially weak password to make it secure (e.g., \cite{forget2008improving,macrae2016strategies,Houshmand}.   
 Our approach with PiXi is entirely different than previous text password nudges; we aim to facilitate the user's password creation without suggesting alternatives, but instead using the following set of nudges immediately prior to password creation: (i) facilitate (positioning and suggesting alternatives) to help users explore an unusual path (and set of selections) through the PiXi system, (ii) confront (throttling mindless activity) to ensure users consider their PiXi selections, and (iii) reinforce (subliminal priming) to make the user's PiXi selections more prominent and easily accessible at the time the user is attempting to conceive a new password.  The goal of this combination of nudges is to create an engaging, interactive, and effective nudge to impact password creation.

\begin{figure*}
  \begin{subfigure}[t]{.5\textwidth}
    \centering
    \tmpframe{\includegraphics[width=0.95\linewidth]{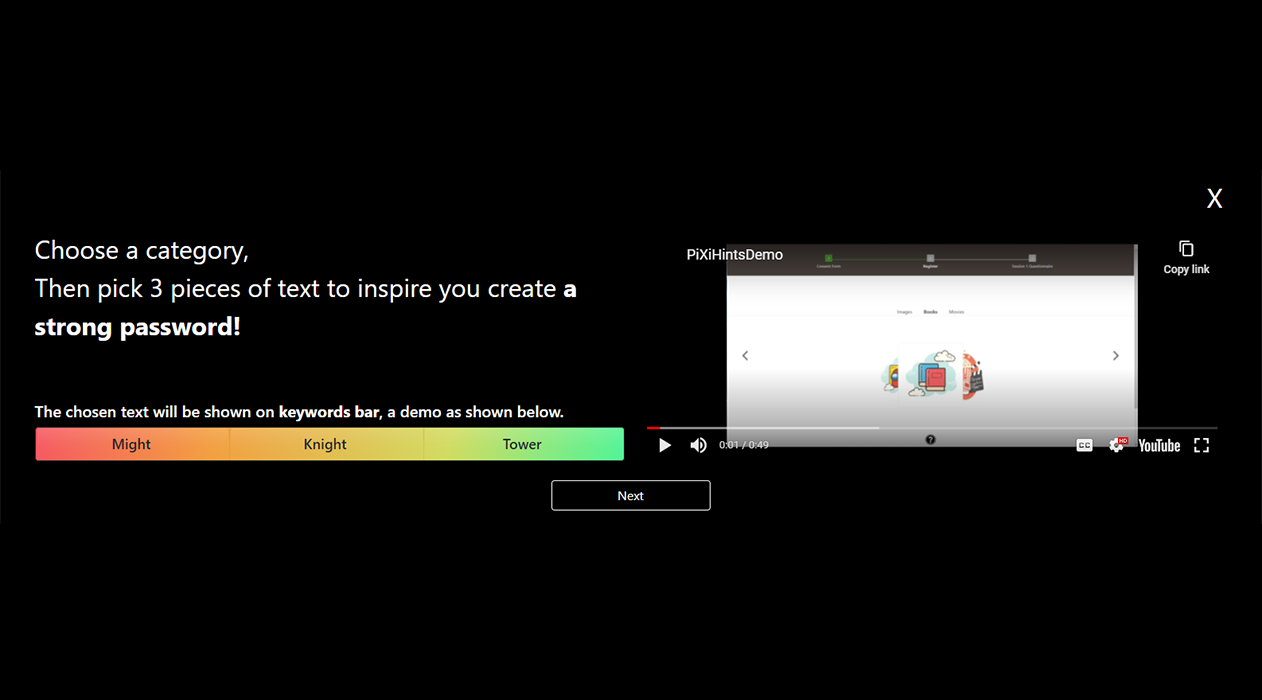}}
    \caption{Introduction Page.}
    \label{fig:FinalModal}
  \end{subfigure}
  \hfill
  \begin{subfigure}[t]{.5\textwidth}
    \centering
    \tmpframe{\includegraphics[width=0.95\linewidth]{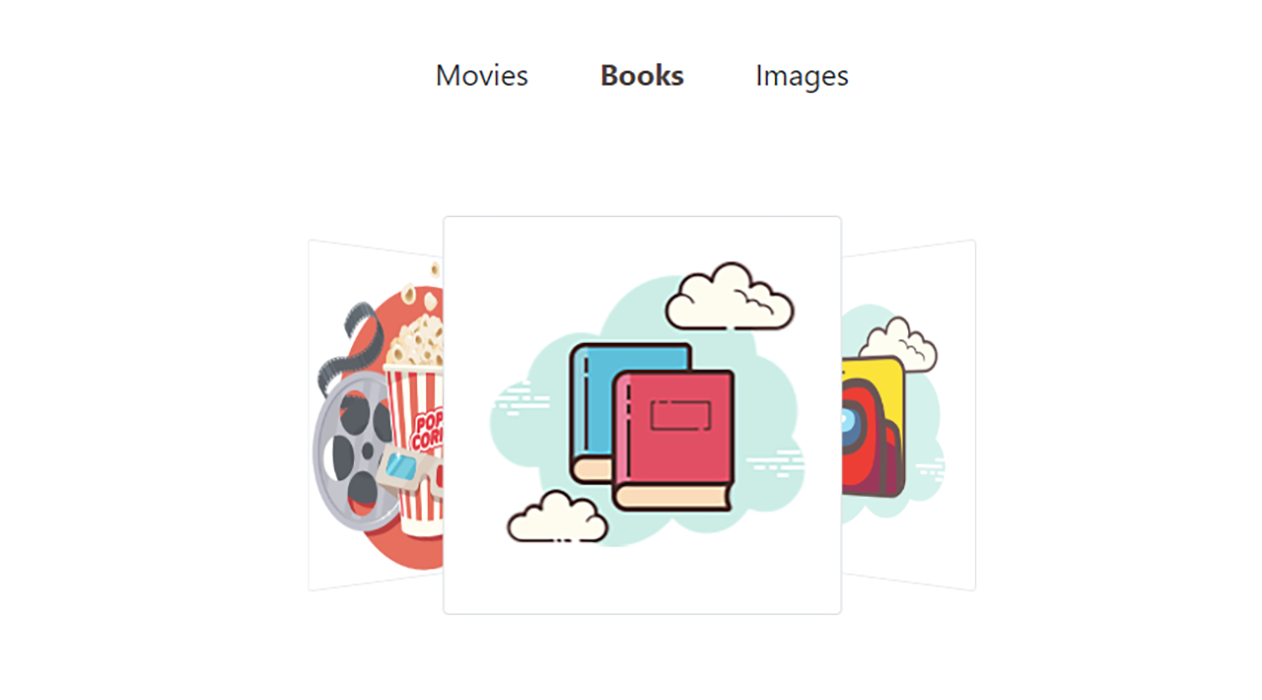}}
    \caption{Category Page.}
    \label{fig:tabs}
  \end{subfigure}
  \medskip
  \begin{subfigure}[t]{.5\textwidth}
    \centering
    \tmpframe{\includegraphics[width=0.95\linewidth]{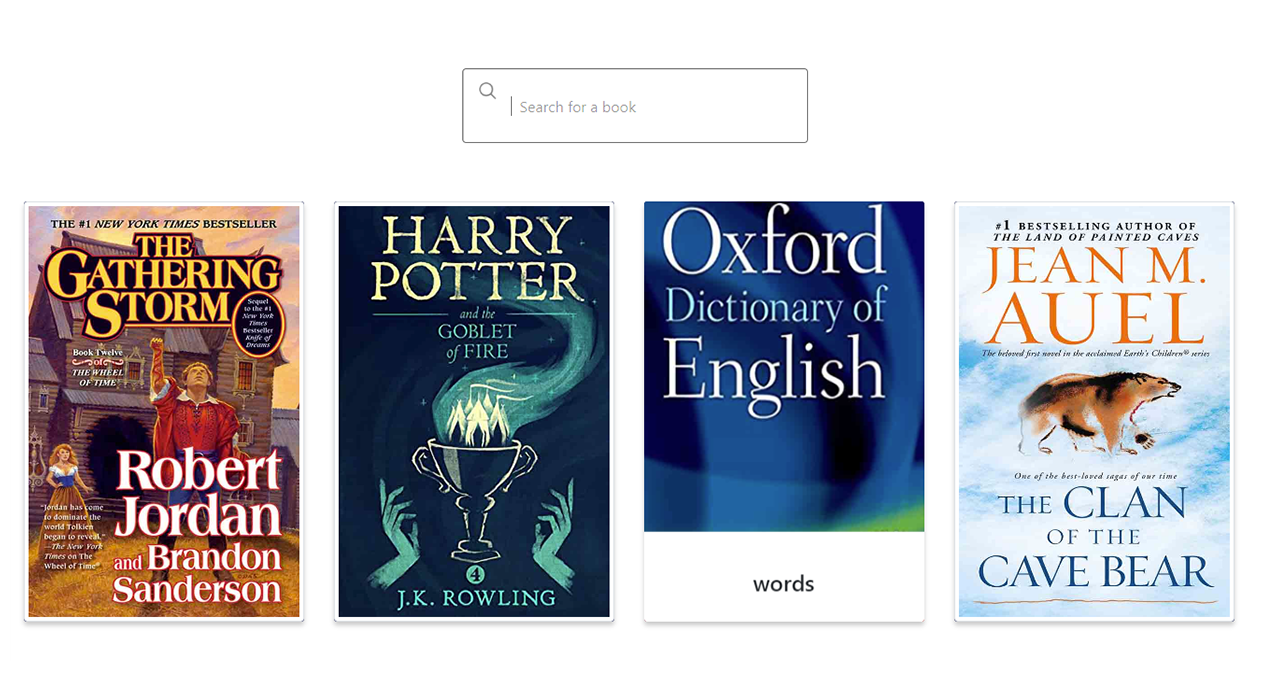}}
    \caption{Item Page, Books.}
    \label{fig:booksTab}
  \end{subfigure}
  \hfill
  \begin{subfigure}[t]{.5\textwidth}
    \centering
    \tmpframe{\includegraphics[width=0.95\linewidth]{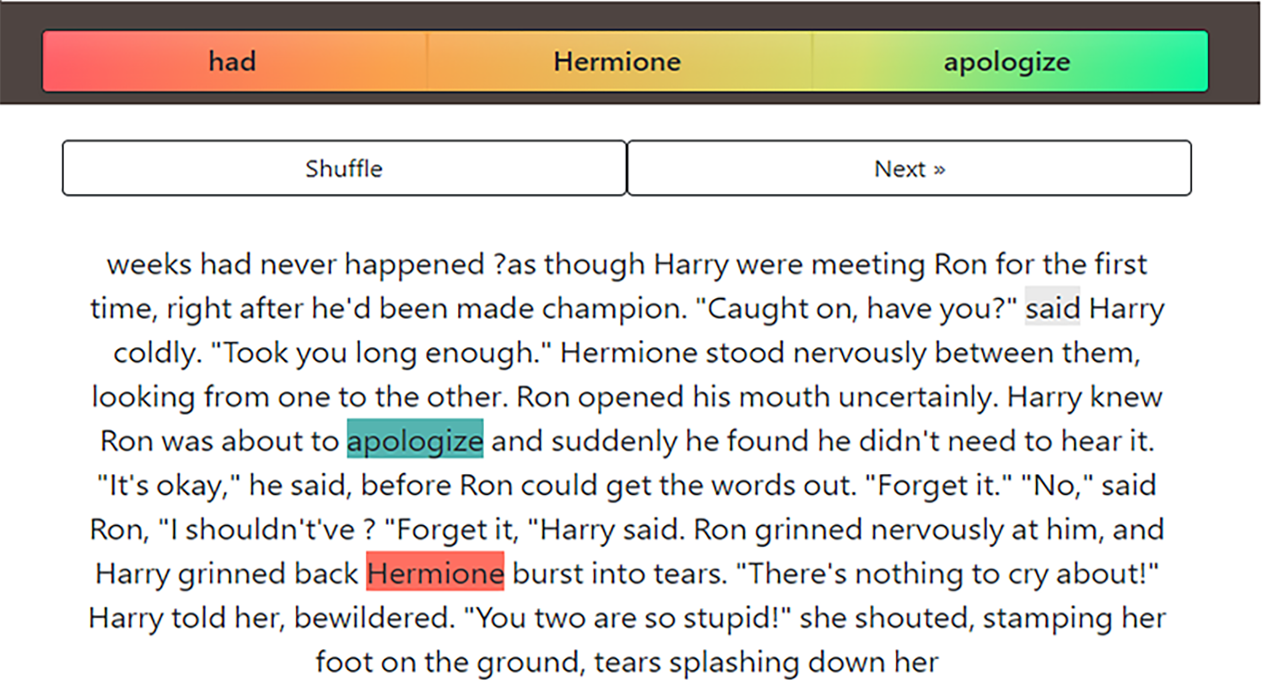}}
    \caption{Keyword Selection Page.}
    \label{fig:keywords}
  \end{subfigure}  
  \medskip
  \begin{subfigure}[t]{.5\textwidth}
    \centering
    \includegraphics[width=0.95\linewidth]{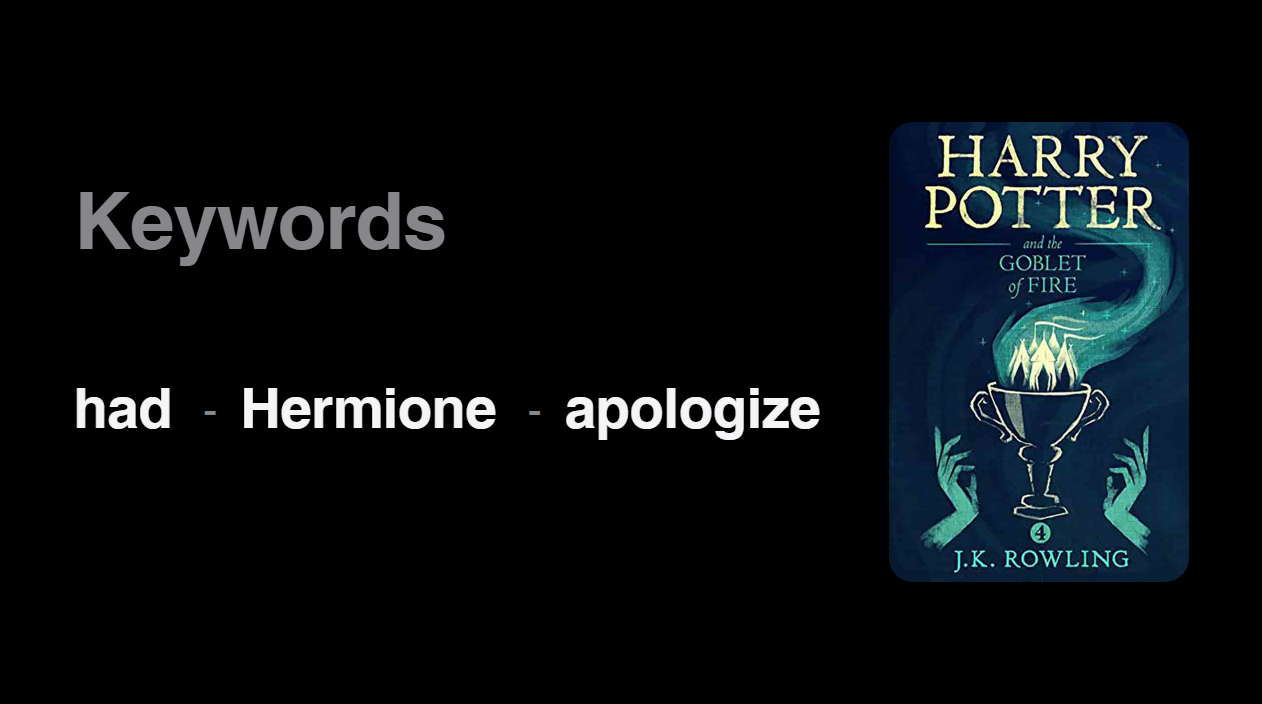}
    \caption{Keyword Splash Page.}
    \label{fig:modal}
  \end{subfigure}  
  \hfill
  \begin{subfigure}[t]{.5\textwidth}
    \centering
    \tmpframe{\includegraphics[width=0.95\linewidth]{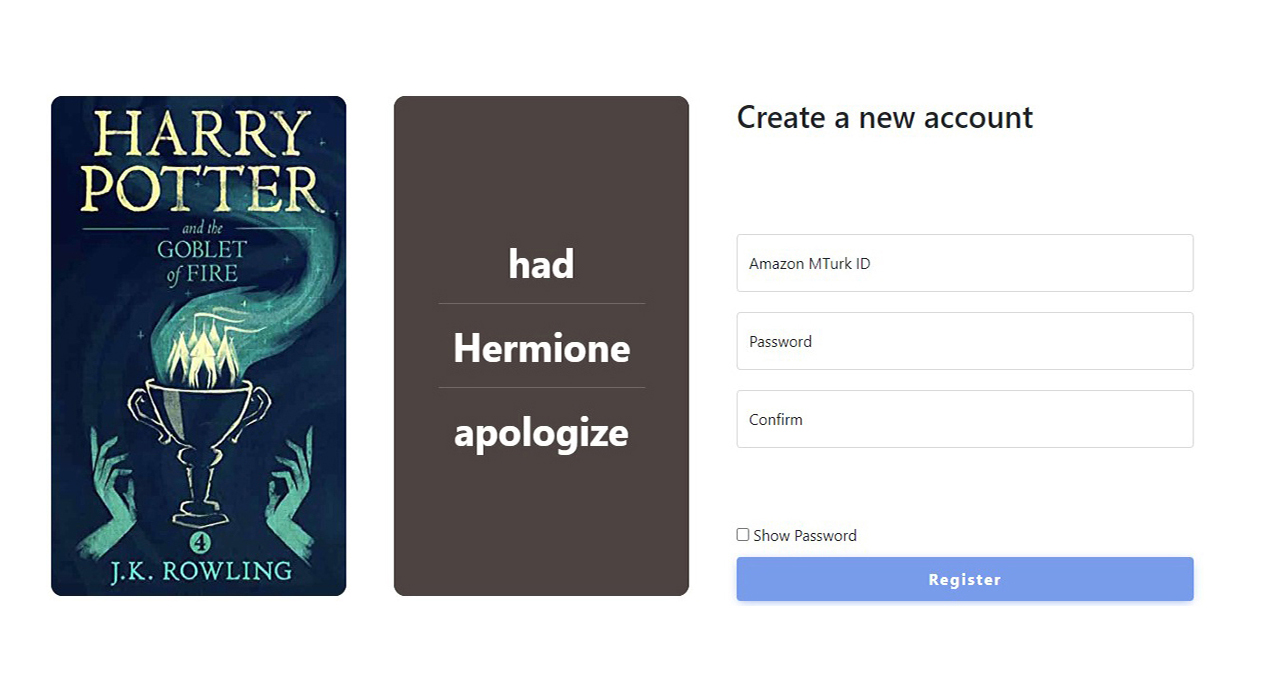}}
    \caption{The Register Page.}
    \label{fig:CPCompare}
  \end{subfigure}     
  \caption{The key user-interaction interfaces in PiXi and its extension PiXi-Hints: (a) the introduction page provides a video tutorial and instructions to users on how to use the system. By clicking the ``Next'' or ``X'' buttons, they will be directed to (b) the category page, which contains three possible content categories: Books, Movies, and Images. Once users select their desired category, they will be taken to (c) the item page, which contains 20 randomly selected items, e.g., book covers in (c). Selecting an item will lead users to (d) the keyword selection page, where they choose three keywords from a random excerpt of the text of the selected item. After selecting all three keywords, users will see the (e) keyword splash page that displays all three chosen keywords (for three seconds) to nudge them further. Finally, users will see (f) the register page which features a large display area of the selected items and keywords on the left side of the typical registration input panel.}
  \label{fig:PIXIComponents}
\end{figure*}
\section{System Design} \label{sd}

The PiXi system aims to nudge users to create stronger passwords, by engaging them with an interactive system for information exploration (e.g., search and select a sequence of keywords) before they create their typical alphanumeric passwords. Instead of limiting user choice, PiXi exposes its users to some unusual and randomized information to shake them out of their typical password creation patterns and get them thinking about new possibilities for their passwords. 

\vskip 1mm
\noindent \textbf{PiXi Components.}
Users interact with PiXi just before password creation through:
\vskip 1mm
\noindent \textit{Introduction.} The introduction page (see Figure \ref{fig:FinalModal}) offers a brief description of the system via a YouTube video tutorial that guides users through the step-by-step process of PiXi. It illustrates how to select a category and a keyword. A short paragraph and a simple animation are also included on the introduction page to assist users in selecting keywords. The users can bypass this page by clicking the ``Next'' or ``X'' buttons, and they can always return to it by clicking on the question icon located at the interface. 

\vskip 1mm
\noindent \textit{Category Selection.} The category page (see Figure \ref{fig:tabs}) contains three possible content categories for user selection: images, books, or movies. The order of categories is randomly shuffled for each user. This page contains a \textit{facilitate (positioning) nudge} \cite{caraban201923} as it positions a category in the center more prominently to nudge the user to select it. The user still has the option to choose another category. Once a category is selected, the user is directed to an item page (see below). 

\vskip 1mm
\noindent \textit{Item Page.} 
The item page contains a set of 20 randomly selected items (e.g., book covers, movie covers, or images) from the selected category.\footnote{Each set of category items is retrieved from its own API. We created the book API and use publicly-available APIs for others.}  
A user then can select an item by clicking its image cover. If not interested in any items, the user can search for her item of interest by the search bar with autocomplete feature. The maximum number of items per page is limited to 20 to maintain an organized user interface. The first row of items, along with the search bar, is shown in Figure \ref{fig:booksTab}. This page contains a \textit{facilitate (suggesting alternatives) nudge} \cite{caraban201923}, by facilitating the selection from a random set of items over many others.
%
%
%
%
\vskip 1mm
\noindent \textit{Keyword Selection Page.} After selecting an item, the user is brought to the keyword selection page, where she must choose three keywords. For example, if a user selects ``Harry Potter 4'' as an item, she will be shown a random excerpt of the book (see Figure \ref{fig:keywords}) from which she is expected to select her keywords. Once each keyword is chosen, it is shown in a bar at the top of the page. We set the maximum number of words per excerpt to 50 to avoid scrolling the page for the user, but the user can click on the shuffle button to land on another random excerpt of the book. After the selection of each keyword, the user is directed to another random excerpt containing the previously selected keyword. Suppose that the user has already selected ``had'' and ``Herminone'' as the first and second keywords. For the third keyword selection, she would be shown a random excerpt containing the word ``Herminone'' (highlighted in red); see Figure \ref{fig:keywords} for this exact scenario. Then, she can select ``apologize'' (highlighted in blue once selected) as the final third keyword. 


\vskip 1mm
\noindent \textit{Keyword Splash.} Once three keywords are selected, the user will be shown the her selected keywords in a ``splash'' page as shown in Figure \ref{fig:modal}. This page intends to employ further nudging towards selected keywords just before the password creation phase. This page has a black background with soft-white text to create a dramatic color contrast for drawing visual attention to selected keywords, and it automatically close after 3 seconds. But users can manually close it by clicking anywhere on the screen. This splash page aims to offer \textit{a confront nudge} (throttling mindless activity) \cite{caraban201923}, to nudge users to review the content again.

\vskip 1mm
\noindent \textit{Registration.} PiXi adds a large display area of selected items and keywords on the left side of the typical registration input panel (see Figure \ref{fig:CPCompare}). This addition serves \textit{a reinforce nudge (or subliminal priming)} \cite{caraban201923}, as they make the image cover and keywords more prominent and easily accessible at the time the user is attempting to conceive a new password. We implement the password length requirement of at least 8 characters. 

\vskip 1mm
\noindent \textit{Login.} PiXi does not modify the standard login page, and users simply need to enter their username and password to complete the login process. 
\vskip 2mm
\noindent \textbf{An Extension: PiXi-Hints.} \label{subsec:pixihints}
PiXi can also be deployed as a hint for password recovery. To this end, we also have designed a PiXi extension called \emph{PiXi-Hints}, which has all the components of PiXi but slightly differs at the login time. It requires the user to interact with PiXi just before login by inputting their keywords. This interaction intends to help users remember their passwords.  In our implementation, we did not require users to recall their keywords but recorded their recall for analysis purposes.\footnote{The introduction video had some minor differences for users of PiXi-Hints: they have an additional sentence that advises them to select interesting and memorable keywords. This recommendation is provided to encourage users to remember their keywords as they will need to reuse PiXi to input them again before each login.}
\section{User Studies} \label{us}
We conducted a two-session study on Amazon MTurk to evaluate PiXi's ability to nudge users. Our study was approved by our university's Research Ethics Board. 

\vskip 1mm
\noindent \textbf{Recruitment and Compensation.}
Our advertisement was made visible to all MTurk workers, but only US workers with an approval rate of 95\% or above were allowed to participate. The users first reviewed and signed the consent form, then were redirected to the PiXi system. 
The sessions were compensated at the US minimum wage at the time of the study (\$7.25/hour). For Sessions 1 and 2 (resp.) with estimated completion times of 7 and 2 minutes (resp.), the participants received \$0.85 and  \$0.35 (resp.).   
\vskip 1mm
\noindent \textbf{Conditions/Groups.}
Upon beginning the study, users were randomly assigned to one of three groups: 
\begin{enumerate}
    \vspace{-5pt}
    \item \textit{Control:} Users create a password and log in as usual (without PiXi).
    \item \textit{PiXi:} Users are asked to use PiXi only prior to password creation. 
    \item \textit{PiXi-Hints:} Users are asked to use PiXi-Hints, which includes using PiXi for both password creation and login.
    \vspace{-5pt}
\end{enumerate}
\vskip 0.1mm
\noindent \textbf{Sessions and Tasks.} Our study contains two sessions. For Session 1, participants were required to register an online account (the process differs based on condition/group), and then complete Questionnaire 1. For Session 2 (7 days later), participants who successfully completed Session 1 were invited back through Amazon MTurk to login to their accounts. After successful login or three unsuccessful login attempts, the participants filled out the exit Questionnaire 2. 
%
%
\vskip 1mm
\noindent \textbf{Data Cleaning.} \label{dataValidation}
To maintain the integrity of our analyses, we have rigorously cleaned our data to remove noisy and unreliable instances. For our analyses, we have removed the data of these participants: (1) \textit{weakly-committed (N=67):}  the users with weak predictable passwords (e.g., MTurk IDs or simple number sequences) or inconsistent responses to the SUS scale's Likert questions ; (2) \textit{multi-identity (N=193)}: the users who participated in our study with multiple accounts or bots\footnote{These 193 participants chose an identical but uncommon password, possibly due to these accounts all being controlled by one.}; and (3) \textit{inattentive (N=58):} users who failed our Likert-scale attention question of ``Seven plus three equals eight'' in Session 1. Overall, we were surprised by the initial amount of noise in our dataset. The final breakdown of user distribution and removal can be found in Table \ref{tab:sessions-completion-rates}. 
\begin{table*}[tb]
\centering
\begin{tabular}{l>{\centering}p{2.6cm}>{\centering}p{2.6cm}>{\centering\arraybackslash}c}
\toprule
\textbf{}                                                            & \textbf{Control} & \textbf{PiXi} & \textbf{PiXi-Hints}\\
\cmidrule(){2-4}
Participants                                      & 181               & 185            & 192 \\
\midrule
Weakly-Committed & 19               & 14             & 34                  \\
Multi-Identity & 76               & 53            & 64                  \\ 
Inattentive & 15               & 35             & 8                  \\ \midrule
Valid Participants (Session 1)                          & 71               & 83            & 84                  \\ 
Valid Participants (Session 2)                           & 10               & 9            & 12                \\ \bottomrule
\end{tabular}
\caption{Statistic of session completion and filtered participants across conditions.}
\label{tab:sessions-completion-rates}
\end{table*}

\vskip 1mm
\noindent \textbf{Demographics} \label{Demographics}
Table \ref{tab:user-demographics} presents an overview of the participant demographics for our study collected through the questionnaire in Session 1. Overall, our participants were composed of 41\% female, 58\% male, and 0.1\% who preferred not to specify their gender. The majority of participants (51\%) fell within the 20–30 age group, followed by the age group of 30–40 making up 32\% of participants. Regarding participants’ education level, most participants (68\%) had a Bachelor’s degree, followed by a Master's degree (25\%). The majority of participants in our study worked in Business (24\%), Technology (21\%), or Health (13\%).
\begin{table*}[h]
\begin{tabular}{p{1.58cm}>{\centering}p{1.1cm}>{\centering}p{1.2cm}>{\centering}p{1.55cm}p{2.12cm}>{\centering}p{1.1cm}>{\centering}p{1.2cm}>{\centering\arraybackslash}p{1.45cm}}
                  \cmidrule[0.6pt](r){1-4} \cmidrule[0.6pt](){5-8}
   \textbf{Gender}&\textbf{Control} &\textbf{PiXi} & \textbf{PiXi-Hints} &
   \textbf{Language}& \textbf{Control} & \textbf{PiXi} & \textbf{PiXi-Hints} \\ 
Female      & 42.3\% & 39.8\% & 40.5\% &  English & 98.6\% & 100.0\% & 98.8\% \\
Male        & 56.3\% & 59\%   & 59.5\% & Other           & 1.4\%     & 0.0\%  &1.2\%     \\
N/A         & 1.4\% & 1.2\%      & 0.0\% & N/A             & 0.0\%     & 0.0\%     & 0.0\%     \\
\cmidrule[0.6pt](r){1-4} \cmidrule[0.6pt](){5-8}
                   \textbf{Age}&\textbf{Control} &\textbf{PiXi} & \textbf{PiXi-Hints} &
   \textbf{Occupation}& \textbf{Control} & \textbf{PiXi} & \textbf{PiXi-Hints} \\ 
   
                   Under 20    & 0.0\%    & 0.0\%      & 0.0\%  & Engineering     & 7.0\%    & 6.0\%     & 7.1\%     \\
                    20-30       & 54.9\%   & 50.6\%   & 48.8\%                                          & Arts and Entmt. & 1.4\%     & 4.8\%     & 7.1\%     \\
                   30-40       & 25.4\%   & 27.7\%   & 34.5\%                                          & Business        & 31.0\%    & 18.1\%  & 26.2\%    \\
                   40-50       & 11.3\%   & 9.6\%   & 9.5\%                                             & Communications  & 4.2\%     & 2.4\%     & 3.6\%     \\
                50-60       & 5.6\%    & 6.0\%    & 6.0\%                                            & Social services & 5.6\%     & 6.0\%   & 2.4\%     \\
                60+         & 2.8\%    & 6.0\%      & 1.2\%      & Education       & 7.0\%    & 7.2\%     & 8.3\%     \\
                   N/A         & 0.0\%    & 0.0\%      & 0.0\%                                            & Technology      & 14.1\%    & 24.1\%  & 23.8\%    \\ 
                   \cmidrule[0.6pt](r){1-4}
                   \textbf{Education}&\textbf{Control} &\textbf{PiXi} & \textbf{PiXi-Hints}                                            & General Labour  & 2.8\%     & 7.2\%   & 1.2\%     \\
                   None        & 0.0\%   & 0.0\%      & 0.0\%                                              & Agriculture     & 1.4\%     & 3.6\%     & 3.6\%     \\
High School & 1.4\%   & 4.8\%   & 8.3\%     & Government      & 2.8\%     & 2.4\%     & 2.4\%     \\
                   Bachelor's  & 74.6\%   & 68.7\%   & 63.1\%                                          & Health          & 18.3\%    & 10.8\%   & 11.9\%     \\
                   Master's    & 23.9\%   & 24.1\%   & 27.4\%         &                                   Law             & 0.0\%    & 0.0\%     & 0.0\%     \\
                   PhD.        & 0.0\%    & 2.4\%      & 1.2\%                                                    & Sales           & 2.8\%     & 4.8\%  & 0.0\%     \\
                   N/A         & 0.0\%    & 0.0\%      & 0.0\%                                                 & N/A             & 1.4\%     & 2.4\%   & 2.4\% \\
                   \cmidrule[0.6pt](r){1-4} \cmidrule[0.6pt](){5-8}
\end{tabular}
\caption{The user demographics across the three conditions.}
\label{tab:user-demographics}
\end{table*}

\section{Results}
We begin by evaluating indicators that PiXi's nudges work in Section \ref{subsec:nudginganalysis}.  We perform an extensive security analysis in Section \ref{SecurityAnalysis}, and usability analysis in Section \ref{UsabilityAnalysis}.

\subsection{Evaluation of Nudging Efficacy} \label{subsec:nudginganalysis}

\begin{table*}[h]
\centering
\begin{tabular}{l>{\centering}p{3.1cm}>{\centering\arraybackslash}p{4.6cm}}
\toprule
 & \textbf{Positioning Nudge (Category Page)} & \textbf{Suggesting Alternatives Nudge (Items Page)}  \\ 
\midrule
\textbf{Books} & 20/56 (35.71\%) & 40/41 (97.56\%)  \\
\textbf{Movies} & 29/59 (49.15\%) & 40/55 (72.73\%)  \\
\textbf{Images} & 30/51 (58.82\%) & 63/71 (88.73\%)  \\ 
\bottomrule
\end{tabular}
\caption{The acceptance rates of the facilitate nudges, combining PiXi and PiXi-Hints.}
\label{tab:defaultI}
\end{table*}


Through various metrics, we evaluate the efficacy of (i) the positioning nudge in the Category Page, (ii) the suggesting alternatives nudge in the Items Page, and (iii) PiXi's overall nudge ability in the users' password. 

\vskip 1mm
\noindent \textbf{Positioning Nudge in Category Page.} Table \ref{tab:defaultI} shows the acceptance rates of the positioning nudge for categories where one category is initially positioned in the center of the Category Page (for both PiXi and PiXi-Hints). Approximately half of the participants accepted the category positioned such that it can be selected without scrolling. There appears to be a slightly higher preference for the Image and Movie categories. 

\vskip 1mm
\noindent \textbf{Suggesting Alternative Nudge in Items Page.} Table \ref{tab:defaultI} also shows the acceptance rates of the suggested alternative nudge in item pages, where the set of 20 randomly selected items initially appeared on the page for both PiXi and PiXi-Hints.  Most users (72\%-97\%, depending on category) accepted one of the suggested items, indicating that this nudge was successful at nudging users towards exploring unique items they might not otherwise consider.  


\begin{table}[tb]
\centering
\begin{tabular}{l>{\centering}p{2.5cm}>{\centering}p{2.5cm}>
{\centering}p{2.5cm}>{\centering\arraybackslash}p{2.5cm}}
\toprule
 & \textbf{1 keyword} & \textbf{2 keywords} & \textbf{3 keywords} & \textbf{Total}\\ 
 \midrule
\textbf{PiXi} & 7 & 12 & 7 & 26/83 (31\%) \\
\textbf{PiXi-Hints} & 11 & 14 & 14 & 39/84 (46\%) \\ 
\textbf{Total} & 22 & 26 & 17 & 65/167 (39\%) \\     
\bottomrule
\end{tabular}
\caption{The keywords usage rate for both PiXi and PiXi-Hints, including direct and indirect use (e.g., uppercase, lowercase, or additional punctuation). }
\label{tab:keywordsUsageRate}
\end{table}

\vskip 1mm
\noindent \textbf{Do PiXi Nudges Influence Resulting Passwords?}
We aim to determine whether PiXi influenced users' password choices. The most straightforward method to measure this is to determine how many users incorporate their keywords directly in their passwords.  
Our findings, shown in Table \ref{tab:keywordsUsageRate}, revealed that 39\% of users (31\% for PiXi, 46\% for PiXi-Hints) incorporated at least one keyword into their passwords. We consider this metric an underestimate of the number of users who are nudged by PiXi, since users may see a relationship between their passwords and keywords that we are unable to detect (e.g., if it is indirectly related and personal in nature). Although it is likely an underestimate, it still provides evidence that a large percentage of users are influenced by the PiXi system during password creation. An emerging critical question is how these nudges have impacted the security of the chosen passwords, which we will address next.

\subsection{Security Analysis} \label{SecurityAnalysis} 
We study the security of passwords created under each condition from different perspectives including their length, ZXCVBN score, and strength against online and offline attacks. We use a significance level of ($\alpha = 0.05$),  and Holm-Bonferroni correction for multiple-comparison correction. This correction performs an adjustment to significance levels when several statistical tests are performed on a single data set. 



\vskip 1mm
\noindent \textbf{Password Length}. We recorded the length of all the passwords as one measure of the password strength of each condition. To determine whether a condition can influence the password length, we test the following Hypothesis:
\begin{itemize}
\vspace{-7pt}
  \item[\textbf{${\mathcal{H}}_{0}$}] \textit{The distribution of password lengths is similar across PiXi, PiXi-Hints, and Control conditions.}
  \item[\textbf{${\mathcal{H}}_{a}$}] \textit{The distribution of password lengths differs between PiXi, PiXi-Hints, and Control conditions.}
  \vspace{-7pt}
\end{itemize}

\noindent The one-way ANOVA test ($df=2$, $N=238$) rejects the null hypothesis ${\mathcal{H}}_{0}$ ($F=6.5, P=0.002$) after Holm-Bonferroni correction (${\alpha^{\prime}_{(1)}} = 0.0167$), indicating a significant difference in password length among the three conditions with a large effect size ($\eta^2 =0.44$). Table \ref{tab:3in1} shows the mean password length for each condition. The Control condition ($\mu=9.35$) had a significantly lower password length compared to PiXi ($\mu=10.87$) and PiXi-Hints ($\mu=11.42$), while the mean in PiXi and PiXi-Hints are comparable. This suggests that PiXi and PiXi-Hints users tend to create longer passwords than those in the Control condition, which can offer security advantages.

\vskip 1mm
\noindent \textbf{Password Score and Strength.} We use the well-known ZXCVBN \cite{zxcvbn} score to measure the strength of each created password for each condition. To determine whether a condition can influence the password score, we test the following hypothesis: 
\begin{itemize}
\vspace{-5pt}
  \item[\textbf{${\mathcal{H}}_{0}$}] \textit{The distribution of ZXCVBN scores is similar across PiXi, PiXi-Hints, and Control conditions.}
  \item[\textbf{${\mathcal{H}}_{a}$}] \textit{The distribution of ZXCVBN scores differs between PiXi, PiXi-Hints, and Control conditions.}
  \vspace{-5pt}
\end{itemize}
%
%
%
\noindent A one-way ANOVA test ($df=2, N=238$) revealed a significant difference in password score among the three conditions ($F=3.868, P=0.022$) with a medium effect size ($\eta^2 = 0.032$), leading us to reject the null hypothesis ${\mathcal{H}}_{0}$ after Holm-Bonferroni correction ($ {\alpha^{\prime}_{(3)}} = 0.05$). As shown in Table \ref{tab:3in1}, the Control condition with an average of ($\mu=1.83$) has a lower password score than PiXi ($\mu=2.16$) and PiXi-Hints ($\mu=2.31$). These findings suggest that passwords created through PiXi and PiXi-Hints are stronger than those created by users in the Control condition.
\begin{table*}[tb]
\centering
\begin{tabular}{l>{\centering}p{2.8cm}>{\centering}p{2.8cm}>{\centering\arraybackslash}p{2.8cm}}
\toprule
 & \textbf{Password Length} & \textbf{ZXCVBN Score} & \textbf{SUS Score} \\ 
\midrule
\textbf{Control} &~~9.35$\pm$1.73&1.83$\pm$1.04 & 56.60$\pm$13.28 \\
\textbf{PiXi} &10.87$\pm$4.38& 2.16$\pm$1.02 & 54.48$\pm$11.93 \\
\textbf{PiXi-Hints} &11.42$\pm$4.01& 2.31$\pm$1.17 & 56.68$\pm$11.49\\ 
\bottomrule
\end{tabular}
\caption{The Mean $\pm$ Std. for password length, password score, and SUS score.}
\label{tab:3in1}
\end{table*}
%
%
%

We evaluate password strength by CMU's Password Guessability Service (PGS) \cite{ur2015measuring} which uses numerous state-of-the-art password cracking algorithms to calculate guessability.\footnote{We also study them by CKL\_PSM--a password strength meter based on the chunk-level PCFG model (CKL\_PCFG). However, the results were quantitatively and qualitatively very similar, thus we don't report them here due to space constraints.} To assess password strength under online and offline attacks, we employed online and offline attack thresholds of $10^{6}$ and $10^{14}$ guesses \cite{Pushing_on_String}.  When a password can be guessed before the online (or offline) attack threshold, we call it \emph{online-unsafe} (or \emph{offline-unsafe}). The summary of our analyses is reported in Table \ref{tab:CMU_Guesses}. Passwords that can withstand offline attacks in PiXi (14.4\%) and PiXi-Hints (32.1\%) are significantly higher than in the Control (7\%) condition. Conversely, weak passwords are more common in the Control (18.3\%) than in PiXi (or 10.8\%) and PiXi-Hints (15.5\%). We conducted a test to determine whether password strength depends on different conditions, by testing the following hypotheses:
%

\begin{itemize}
  \item[\textbf{${\mathcal{H}}_{0}$}] \textit{The distribution of password strength measurements is similar across PiXi, PiXi-Hints, and Control conditions.}
  \item[\textbf{${\mathcal{H}}_{a}$}] \textit{The distribution of password strength measurements differs between PiXi, PiXi-Hints, and Control conditions.}
\end{itemize}

We performed a ${\chi^2}$ test ($df=4, N=238$) to examine these hypotheses. The results in Table \ref{tab:CMU_Guesses} showed a significant difference ($\chi^2=17.120$, $P=0.002$) with a medium effect size (Cramer’s $V = 0.187$) across different conditions, so we reject the null hypothesis (${\mathcal{H}}_{0}$) after Holm-Bonferroni correction ($ {\alpha^{\prime}_{(2)}} = 0.025$). 
This finding further supports that PiXi and PiXi-Hints encourage users to create more unique and stronger passwords than the Control condition.


\begin{table}[tb]
\centering
\begin{tabular}{l>{\centering}p{2.8cm}>{\centering}p{2.8cm}>{\centering\arraybackslash}p{2.8cm}}
\toprule
 & \textbf{Online-unsafe} & \textbf{Offline-unsafe} & \textbf{Safe} \\ 
 \midrule
\textbf{Control} & 18.3\% & 74.7\% & 7\% \\
\textbf{PiXi} & 10.8\% & 74.8\% & 14.4\% \\
\textbf{PiXi-Hints} & 15.5\% & 52.4\% & 32.1\% \\ 
\bottomrule
\end{tabular}
\caption{Passwords guessability at the online and offline thresholds of $10^6$ and $10^{14}$, CMU's Password Guessability Service.}
\label{tab:CMU_Guesses}
\end{table}

\vskip 1mm
\noindent \textbf{Should Users Incorporate Keywords in Passwords?} \label{PasswordHintsSimilarity}
As observed in Section \ref{subsec:nudginganalysis}, many users incorporate their keywords into their passwords.  Here we aim to determine the security impact of this behavior, to determine whether PiXi should encourage or prevent it.   As shown in Table \ref{tab:keywordsPasswordsDetails}, for both PiXi and PiXi-Hints, the passwords using keywords had much higher length, score, and guesses than the average passwords. This suggests that users who used keywords were able to create stronger and longer passwords, and as such future versions of PiXi might encourage this behavior. 

\vskip 1mm
\noindent \textbf{Do Some Categories Nudge Stronger Passwords?}
We also investigate whether password strength depends on the nudge category (Books, Movies, or Images). Table \ref{tab:CMU_cate} shows that passwords created by users who selected Books were most resistant to online and offline attacks. Passwords created by users who selected Images have the least ``safe" passwords.  One possible reason for this is that keywords from the Images category tend to be less unique compared to the other categories. These results suggest that password strength differs between categories and that future PiXi implementations might avoid using the Images category.

\begin{table}[tb]
\centering
\begin{tabular}{l>{\centering}p{2.5cm}>{\centering}p{2.5cm}>
{\centering}p{2.5cm}>{\centering\arraybackslash}p{2.5cm}}
\toprule
 & \textbf{Keywords} & \textbf{Length} & \textbf{Score} & \textbf{CMU Guesses}\\ 
\cmidrule[0.4pt](lr){2-5}
\multirow{2}{*}{\textbf{PiXi}} & Yes & 14.15 & 2.81 &$10^{15.45}$\\
& No & 9.25 & 1.89 &$10^{8.89}$ \\
\midrule
\multirow{2}{*}{\textbf{PiXi-Hints}}& Yes & 13.05 & 2.51 &$10^{14.37}$\\
 & No & 9.79 & 2.17 &$10^{10.61}$\\ 
\bottomrule
\end{tabular}
\caption{Comparison of security metrics for passwords with vs. without keywords.  
}
\label{tab:keywordsPasswordsDetails}
\end{table}

\begin{table}[h]
\centering
\begin{tabular}{l>{\centering}p{2.8cm}>{\centering}p{2.8cm}>{\centering\arraybackslash}p{2.8cm}}
\toprule
 & \textbf{Online-unsafe} & \textbf{Offline-unsafe} & \textbf{Safe} \\ 
 \midrule
\textbf{Books} & 7.3\% & 56.1\% & 36.6\% \\
\textbf{Movies} & 14.5\% & 56.4\% & 29.1\% \\
\textbf{Images} & 15.5\% & 73.2\% & 11.3\% \\ 
\bottomrule
\textbf{Total} & 14.7\% & 66.8\% & 18.5\% \\ \hline
\end{tabular}
\caption{The guessability of Passwords at the online and offline thresholds across three categories, combining PiXi and PiXi-Hints.}
\label{tab:CMU_cate}
\end{table}

\subsection{Usability Analysis} \label{UsabilityAnalysis}
We analyze the usability of PiXi and PiXi-Hints, according to (a) SUS score, (b) user satisfaction, (c) login times, and (d) login rates.  Results suggest that PiXi shows promise; most users agreed that it helped them to choose a secure and memorable password, recall rates were promising, and SUS scores were comparable to the Control group.

\begin{figure}[tb]
\centering
  \includegraphics[width=0.6\textwidth]{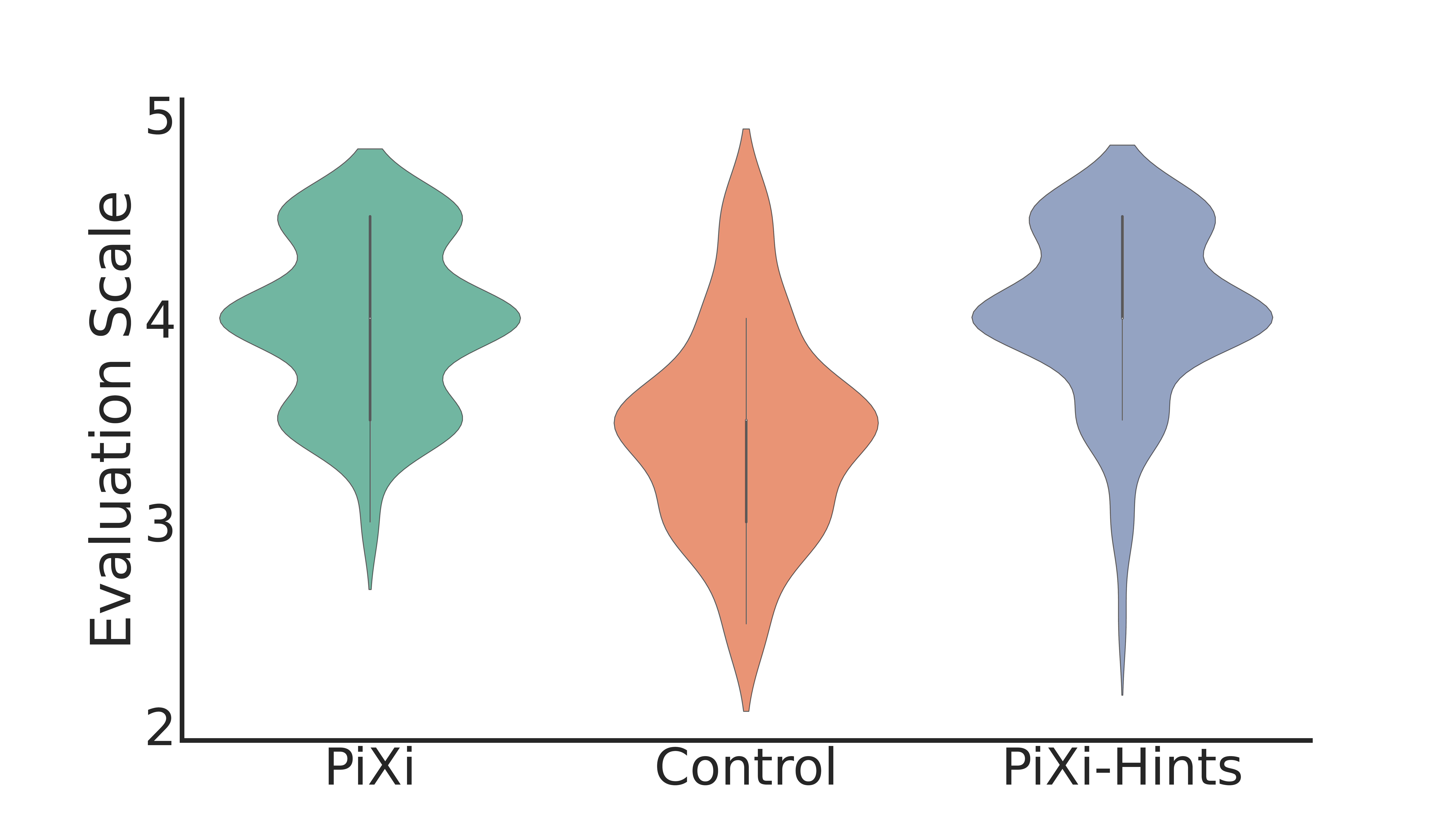} 
  \caption{The violin plot of user satisfaction distributions for three conditions. 
  PiXi and PiXi-Hints users have a similar score distribution, with the majority of users reporting scores of 4 or higher, while Control users have scores concentrated between 3 and 4. }
  \label{Q5:distribution}
\end{figure}

\vskip 1mm
\noindent \textbf{SUS Score.} \label{SUSScore}
To measure the usability of the Control, PiXi and PiXi-Hints, we compare the System Usability Scale (SUS)--- a commonly-used questionnaire to measure the usability of a system \cite{sus1995}. SUS consists of 10 questions that were asked in our Session 1 questionnaire. 
As shown in Table \ref{tab:3in1}, the SUS score is comparable across conditions, supporting that PiXi has no noticeable usability impact. 
%
%
%
%
%
%
\vskip 1mm
\noindent \textbf{User Satisfaction.} \label{Q5usersatisfaction}
To determine the extent to which participants value each password system/process, 
we asked users their level of agreement with the question ``I believe this password creating method helped me to choose a secure and memorable text password.'' 
Figure \ref{Q5:distribution} gives a visual representation of the distribution of the answers, where 5 is for strongly agree, and 1 for strongly disagree.
The users of PiXi or PiXi-Hints (with averages of 3.95 and 4.05) report higher levels of agreement compared to those in the Control condition (with an average of 2.9). Thus, PiXi and PiXi-Hints systems were successful at inspiring/nudging users to select secure and memorable passwords.

﻿
 
﻿
﻿
﻿
﻿


\vskip 1mm
\noindent \textbf{Login Rates and Times.} \label{MemorabilityAnalysis}
We analyze our login data from Session 2 for indications of usability and memorability problems in each condition.  
While the MTurk return rate was low for Session 2, we believe exploring this information can still provide useful insights about system memorability.  
Table \ref{tab:recall} shows the login success rates (over 3 login attempts) and login time.  While the Control group has a higher rate of login failure, we only see this as an indication that PiXi shows promise for helping create stronger and possibly more memorable passwords, and as such further study is required for any concrete statistical analyses.
\begin{table}[tb]
\centering
\begin{tabular}{llll}
\begin{tabular}{l>{\centering}p{2.8cm}>{\centering}p{2.8cm}>{\centering\arraybackslash}p{2.8cm}}
\toprule
 & \textbf{Control} & \textbf{PiXi} & \textbf{PiXi-Hints} \\ 
\midrule
\textbf{Login time} & 14.87$\pm$7.38 & 27.68$\pm$22.1 & 139.5$\pm$36.08 \\
\textbf{Login success rate} & 7/10 (70\%) & 8/9 (88.9\%) & 10/12 (83.3\%) \\
\bottomrule
\end{tabular}
\end{tabular}
\caption{Login data for each condition.}
\label{tab:recall}
\end{table}


As shown in Table \ref{tab:recall}, PiXi-Hints with the additional hint task have higher login times compared to Control. However, surprisingly, PiXi requires a longer login time than Control, while Pixi users tended to require more than one login attempt, which increased the average login time. This issue should be analyzed in future work to determine whether it improves over successive logins or not.   

\section{Conclusion} \label{cl}
We designed, implemented, and studied the efficacy of PiXi (\textbf{P}assword \textbf{i}nspiration by e\textbf{X}ploring \textbf{i}nformation)---a novel approach to nudge users towards creating secure passwords. PiXi is the first approach we are aware of that employs a text password creation nudge that supports users in the task of coming up with a unique password themselves.  PiXi's concept is to ask users to explore unusual information just prior to password creation, to shake users out of their typical habits and thought processes, in the hopes it inspires them to create unique (and therefore stronger) passwords. The results of our study ($N=238$) indicate that PiXi is successful at nudging users to create secure passwords, without explicitly asking them to do so.  Our findings indicate that PiXi users created passwords that are significantly longer and more resistant to password-guessing attacks. PiXi had a comparable overall perception to typical password creation systems, and users agreed that PiXi helped them to create more secure and memorable passwords.  

Our study has some limitations due to Amazon MTurk, which introduced a notable amount of noise in our collected data. While we did our best to fairly catch noise and remove it from our data, it is possible we couldn't catch and filter all noisy data. 
However, since the noise should be consistent between each group, any statistically significant finding should be reliable. Future studies should focus on other populations or enhanced methods to filter noise on MTurk. Such future studies should also focus on long-term recall rates and login times over successive logins. It would also be interesting to study whether a shortened version of the PiXi system (e.g., involving only one keyword) could be equally effective at nudging users toward choosing secure passwords.

Future work also includes designing and evaluating extensions to the PiXi system. PiXi presently only offers three categories. In future work, we suggest the study of additional categories (e.g., music/songs, videos, maps, news, or blog posts) to provide users with a broader set of unique paths/nudges through the system. 

Our results should stimulate future research into both PiXi itself and more generally novel password creation nudges to support users in secure password creation.


\section*{Acknowledgments}
This research was supported by Natural Sciences and Engineering Research Council of Canada (NSERC).
%
%
%
 \bibliographystyle{splncs04}
 \bibliography{references}
%

\end{document}